\documentclass[aps,twocolumn,showpacs,amsmath,amssymb,floatfix]{revtex4-1}

\usepackage{graphicx}
\usepackage{dcolumn}
\usepackage{bm}

\begin{document}

\title{Uncertainties in modeling low-energy neutrino induced reactions on iron group nuclei}



\author{N. Paar$^{1}$, T. Suzuki$^{2}$, M. Honma$^{3}$,T. Marketin$^{1,4}$, D. Vretenar$^{1}$}
\affiliation{$^{1}$Physics Department, Faculty of Science, University of Zagreb, Croatia}

\affiliation{$^{2}$Department of Physics, College of Humanities and Sciences, Nihon University,
Sakurajosui 3-25-40, Setagaya-ku, Tokyo 156-8550, Japan}

\affiliation{$^{3}$Center for Mathematical Sciences, University of Aizu,
Aizu-Wakamatsu, Fukushima 965-8580, Japan}

\affiliation{$^{4}$GSI Helmholtzzentrum f\"{u}r Schwerionenforschung,
Planckstra\ss e 1, D-64291 Darmstadt, Germany}


\begin{abstract}
Charged-current neutrino-nucleus cross sections for $^{54,56}$Fe and $^{58,60}$Ni 
are calculated and compared using frameworks based on relativistic and Skyrme energy density functionals,  
and the shell model. The current theoretical uncertainties in modeling neutrino-nucleus
cross sections are assessed in relation to the predicted Gamow-Teller transition strength and
available data, multipole decomposition of the cross sections, and cross sections averaged 
over the Michel flux and Fermi-Dirac distribution. Employing different microscopic approaches and
models, the DAR neutrino-$^{56}$Fe cross section and its theoretical uncertainty are 
estimated: $<\sigma>_{th}$=(258$\pm$57) $\times$10$^{-42}$cm$^2$, in very good agreement with 
the experimental value:
$<\sigma>_{exp}$=(256$\pm$108$\pm$43) $\times$ 10$^{-42}$cm$^2$. 
\end{abstract}

\maketitle

Weak neutrino-induced processes in nuclei provide information of relevance for modeling
the response in neutrino detectors, understanding the fundamental properties of the weak interaction,
and the role of neutrinos in stellar environment.
Data on neutrino-nucleus cross sections are presently available 
only for $^{12}$C and $^{56}$Fe target nuclei, obtained by the
LSND~\cite{Ath.97} and KARMEN~\cite{Bod.94,Mas.98} 
collaborations, and at LAMPF~\cite{Kra.92}. At present only theoretical approaches can provide 
cross sections for a large number of target nuclei that are involved in various applications of neutrino physics 
and astrophysics. It is, therefore, crucial to quantitatively assess the theoretical uncertainties in
modeling neutrino induced processes, including the detailed structure of principal transitions involved,
and the total cross sections averaged over selected neutrino fluxes. 
The evaluation of current theoretical uncertainties in modeling neutrino induced
processes is also important in view of future experimental programs, e.g., 
spallation neutron source (SNS) at ORNL~\cite{Avi.03}, Large Volume Detector in 
Gran Sasso (LVD)~\cite{Aga.07}, and the beta-beams for the production of neutrinos by 
using the $\beta$-decay of boosted radioactive ions~\cite{Zuc.02,Vol.04}.

Over the years a variety of microscopic models have been developed and
employed in the calculation of neutrino-nucleus cross sections at low energies.
These include the shell model~\cite{Hax.87,Eng.96,Hay.00,Vol.00,Suz.09}, the random phase 
approximation (RPA)~\cite{Aue.97,Sin.98,Vol.00}, 
continuum RPA (CRPA)~\cite{Kol.92,Jac.99,Bot.05},
hybrid models of CRPA and shell model~\cite{Aue.97,KLM.99}, 
Fermi gas model~\cite{Kur.90}, quasiparticle RPA~\cite{Laz.07,Che.10},
projected QRPA~\cite{Sam.11}, and relativistic quasiparticle 
RPA (RQRPA)~\cite{Paar.08}. For the purpose of the present analysis
of theoretical uncertainties in modeling neutrino-nucleus cross sections,
we chose the iron group nuclei for which the framework
based on the energy density functionals and the shell model
represent feasible approaches and, in addition, data from muon 
decay at rest (DAR) are available~\cite{Bod.94,Mas.98}. The framework based on 
energy density functionals employs the Relativistic Hartree-Bogoliubov 
model (RHB) to determine the nuclear ground state, and the RQRPA to calculate 
all relevant transitions induced by the incoming neutrinos~\cite{Paar.08}. Model calculations are
performed using effective interactions with density-dependent meson-nucleon 
couplings, in this case the DD-ME2 interaction~\cite{LNVR.05}, whereas pairing correlations 
are described by the finite range Gogny force~\cite{BGG.91}. The nuclear shell model
employed in the present study  is based on the GXPF1J 
effective interaction~\cite{Hon.05} for $\lambda^{\pi}=1^+$ channel,
supplemented by the RPA based on a Skyrme functional (SGII) for other multipoles~\cite{Suz.09}.
A detailed analysis of (anti)neutrino - $^{56}$Fe cross sections based on the QRPA with  
Skyrme functionals is given in Ref.~\cite{Laz.07}. Shell-model calculations are carried out with the code MSHELL~\cite{Miz.00}.
\begin{figure}
\centerline{
\includegraphics[width=80mm,angle=0]{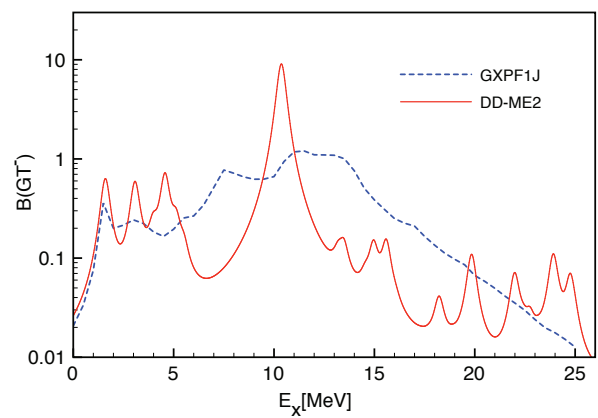}
}
\caption{(Color online) GT$^-$ transition strength in $^{56}$Fe, 
calculated using the  
RQRPA (DD-ME2) and the Shell model (GXPF1J).
}
\label{gt_minus_Fe_Ni.eps}
\end{figure}

Because of its importance for neutrino-nucleus cross sections at low energies, we
start by analyzing the Gamow-Teller(GT) transition strength in the iron group nuclei.
Fig.~\ref{gt_minus_Fe_Ni.eps} displays the GT$^-$ strength distributions 
for $^{56}$Fe, obtained using the shell model (GXPF1J) and the RQRPA (DD-ME2).
In both cases the calculated transition strength is folded by a Lorentzian 
with the width $\Gamma$=0.5 MeV. 
The RQRPA includes only $2qp$ configurations and, therefore, cannot provide 
the detailed structure of excitation spectra obtained by the shell model. 
Nevertheless, one can observe that the calculated transition strength distributions 
are in reasonable agreement.

\begin{figure}
\centerline{
\includegraphics[width=85mm,angle=0]{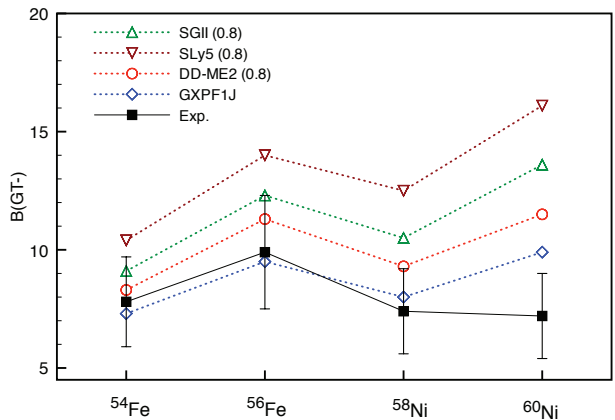}
}
\caption{(Color online) GT$^-$ transition strengths calculated with the Skyrme RPA (SGII, SLy5) 
and RQRPA (DD-ME2), in comparison to the shell model (GXPF1J)~\cite{Suz.09} and experimental 
values~\cite{Rap.83}. (Q)RPA calculations include the quenching factor 
0.8 in the axial-vector coupling constant $g_A$.}
\label{gt_minus_Fe_Ni_q08.eps}
\end{figure}

\begin{figure}
\centerline{
\includegraphics[width=85mm,angle=0]{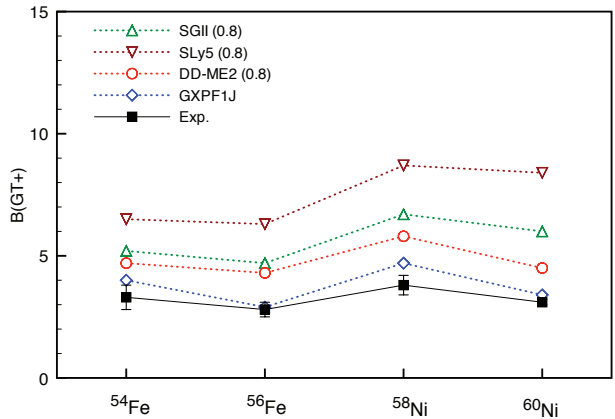}
}
\caption{(Color online) Same as described in the caption to Fig.\ref{gt_minus_Fe_Ni_q08.eps} but for the GT$^+$ transitions.
Experimental values are from Refs.~\cite{Vet.89, Kat.94,Wil.95}.}
\label{gt_plus_Fe_Ni_q08.eps}
\end{figure}

In Figs.~\ref{gt_minus_Fe_Ni_q08.eps} and \ref{gt_plus_Fe_Ni_q08.eps}
the GT$^\pm$ transition strengths contributing to the neutrino-induced processes 
are compared for a set of iron group nuclei: $^ {54,56}$Fe and $^{58,60}$Ni. The following models and respective parameterizations have been used: i) RPA based on Skyrme functionals (SGII, SLy5), ii) RQRPA (DD-ME2)~\cite{LNVR.05}, and iii) shell model (GXPF1J)~\cite{Suz.09}. 
Results of model calculations are compared with the data for GT$^-$~\cite{Rap.83} and 
GT$^+$~\cite{Vet.89, Kat.94,Wil.95} transition strengths. 
The (Q)RPA calculations include the quenching of the free-nucleon axial-vector coupling constant $g_A = 1.262 \to g_A = 1$, corresponding to the quenching factor 0.8 in the GT transition operator. A quenching factor is also used in the shell model. However, its value 0.74 is adapted to the effective interaction and model space under consideration~\cite{Suz.09}. One notices in Figs.~\ref{gt_minus_Fe_Ni_q08.eps}  and \ref{gt_plus_Fe_Ni_q08.eps} that the shell model reproduces the experimental values of GT$^-$ transition
strength with high accuracy (except for $^{60}$Ni), and also the GT$^+$ strength is reasonably reproduced. 
(Q)RPA based approaches, however, even by quenching the value
$g_A \to 1$, overestimate both the GT$^-$ and GT$^+$ transition strengths.
For $^ {54,56}$Fe and $^{58}$Ni the relativistic QRPA results for the B(GT$^-$) are within experimental uncertainties. 
\begin{table}
\caption{GT$^-$ transition strengths calculated using the Skyrme RPA (SGII, SLy5), 
RQRPA (DD-ME2), and shell model (GXPF1J)~\cite{Suz.09} (all using $g_A^*=0.74g_A$),  compared to experimental values~\cite{Rap.83}. }
\begin{center}
\item[]\begin{tabular}{@{}*{5}{l}}
\hline                        
$ $&$^{54}$Fe&$^{56}$Fe&$^{58}$Ni&$^{60}$Ni \cr
\hline
SGII & 7.8&10.5 & 9.0&  11.7\cr 
SLy5&8.9 &11.9 & 10.7  & 13.8\cr
DD-ME2 &7.1 & 9.7&7.9 &9.8 \cr
GXPF1J &7.3 & 9.5 & 8.0&9.9\cr
\hline
Exp. &7.8$\pm$1.9 &9.9$\pm$2.4 &7.4$\pm$1.8 & 7.2$\pm$1.8\cr
\hline
\end{tabular}
\end{center}
\label{table1}
\end{table}
%
The fact that QRPA calculations systematically overestimate the
measured GT$^{\pm}$ transition strength, even though different effective interactions are used, 
indicates that a somewhat stronger quenching of the axial-vector coupling constant might be necessary 
to reproduce the data. Actually, if the same quenching factor 0.74 used by the shell model is also employed
in (Q)RPA calculations, a very good agreement with the shell-model results is obtained.
The (Q)RPA and shell model B(GT$^-$) values, all obtained using the same quenching 
factor 0.74, are shown in Table I in comparison to the data~\cite{Rap.83}. 
In this case for all four nuclei: $^ {54,56}$Fe and $^{58,60}$Ni, the (Q)RPA B(GT$^-$) values 
are found in good agreement with the shell-model results, particularly for the 
relativistic QRPA (DD-ME2). We have verified that the same result is also obtained for the B(GT$^+$) 
channel. This similarity is not obvious as the two theoretical frameworks have 
different foundations, use different effective interactions
and model spaces, that is, the underlying structure cannot be directly compared, except for
the resulting observables that one confronts to data.
This result should also be considered with caution because of well known problem of missing GT 
strength, due either to excitations that involve complex configurations at higher excitation
energies, or excitations that include non-nucleonic degrees of freedom.

As already emphasized in previous studies of neutrino-nucleus reactions~\cite{Laz.07,Paar.08},  
not only GT$^-$ transitions but also excitations of higher multipoles must be included, 
depending on the energy range under consideration. In this work we analyze  
the reaction $^{56}$Fe($\nu_e,e^-$)$^{56}$Co in more detail, using two representative theoretical approaches: 
the shell model (GXPF1J) supplemented by the RPA based on Skyrme functionals (SGII), and the 
fully consistent relativistic framework RHB+QRPA (DD-ME2). The goal is to provide an estimate
of theoretical uncertainties of contributions from different multipole transitions to the 
neutrino cross section.
\begin{figure}
\centerline{
\includegraphics[width=85mm,angle=0]{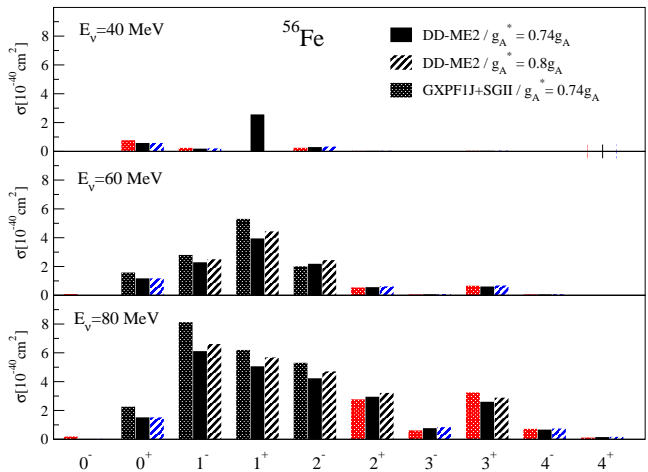}
}
\caption{(Color online) Contributions of the multipole transitions $\lambda^{\pi}=0^{\pm} - 4^{\pm}$ to the 
inclusive cross section for the $^{56}$Fe($\nu_e,e^-$)$^{56}$Co reaction,
at $E_{\nu_e}$ = 40, 60 and 80 MeV.
Calculations include RHB+RQRPA (DD-ME2), and the shell model (GXPF1J) (for the $1^+$ 
transition) plus the RPA (SGII) for higher multipoles. The quenching factors in $g_A$ are
denoted in the figure. 
}
\label{FE56_shell_vs_rpa.eps}
\end{figure}
In Fig.~\ref{FE56_shell_vs_rpa.eps} we plot the contributions of the 
multipole transitions $\lambda^{\pi}=0^{\pm} - 4^{\pm}$ to the 
inclusive cross section for the $^{56}$Fe($\nu_e,e^-$)$^{56}$Co reaction,
at $E_{\nu_e}$ = 40, 60 and 80 MeV. 
The RHB+RQRPA (DD-ME2) calculations include the standard (0.8), and enhanced (0.74) 
quenching factors of the axial-vector coupling constant $g_A$  in all multipole operators. 
In the non-relativistic framework the shell model (GXPF1J) is used for $1^+$ transitions, and 
RPA (SGII) is used for higher multipoles. The quenching 
factor for the axial-vector coupling $g_A$ is 0.74. 
In the shell model calculation of 1$^{+}$ transitions the effect of finite momentum transfer ($q$) is 
taken into account by evaluating the matrix elements $<f || j_0(qr)[Y^0\times \vec{\sigma}]^{1}t_{-} ||i>$ and  $<f || j_2(qr)[Y^2\times \vec{\sigma}]^{1}t_{-} ||i>$ at each $q$, instead of the approximate treatment of Ref. \cite{Suz.09} where the GT matrix element $<f|| j_0(qR)\vec{\sigma}t_{-} ||i>$ was evaluated multiplied by $j_0(qR)$ (R is the nuclear radius).

At relatively low neutrino energies ($E_{\nu} \lesssim$ 40 MeV) the dominant contribution to the 
calculated cross sections originates from GT transitions ($\lambda^{\pi}$=1$^+$). With increasing 
$E_{\nu}$, however, contributions from other multipole transitions become important. 
In particular, at $E_{\nu}$=80 MeV the dominant transition is the spin-dipole 
$\lambda^{\pi}$=1$^-$, but also other components, e.g. $\lambda^{\pi}$=1$^+$,  2$^-$, 2$^+$, 3$^+$, 
play an important role. The cross sections plotted in Fig.~\ref{FE56_shell_vs_rpa.eps} show that the 
two models predict a very similar structure and distribution of the relative contributions from various 
multipoles. These results are also in agreement with those discussed in Ref.~\cite{Laz.07}. 

\begin{figure}
\centerline{
\includegraphics[width=85mm,angle=0]{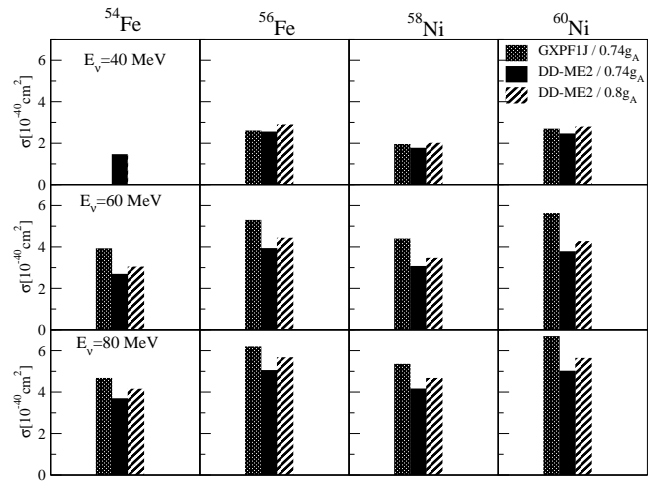}
}
\caption{(Color online) Inclusive neutrino-nucleus cross sections in the $1^+$ channel 
for the $^{54,56}$Fe and $^{58,60}$Ni target nuclei, at the incoming neutrino energies 
$E_{\nu_e}$ = 40, 60 and 80 MeV. The results are obtained
using the RHB+RQRPA (DD-ME2) framework, and the shell model (GXPF1J). 
The axial-vector coupling $g_A^*$ 
includes a quenching factor as denoted in the figure.
}
\label{FE56_multipoles}
\end{figure}

In Fig.~\ref{FE56_multipoles} we compare the neutrino-capture cross 
sections for the $\lambda^{\pi}$=1$^+$ channel on the set of
target nuclei: $^ {54,56}$Fe and $^{58,60}$Ni, and for the incoming neutrino energies 
$E_{\nu_e}$ = 40, 60 and 80 MeV. The results are obtained
using the RHB+RQRPA (DD-ME2) and the shell model (GXPF1J). 
The axial-vector coupling $g_A^*$ includes a quenching factor as denoted in the figure.
The cross sections increase in heavier isotopes because electron neutrinos are captured 
on neutrons. The two models, although based on 
different microscopic pictures, predict rather similar cross sections. One notes that at 
higher neutrino energies the shell model cross sections are slightly larger than those 
calculated with the RQRPA. Related to the previous discussion on the overall GT strength and 
the quenching of $g_A$, from the cross sections shown in Fig.~\ref{FE56_multipoles} 
it appears that the (Q)RPA does not require a stronger quenching than the usual
$g_A=1$  to be in agreement with the shell model. The reason is that
the calculated cross sections are not only determined by the overall GT transition strength, 
but also by the transition energies that govern the energies of outgoing electrons.

\begin{figure}
\centerline{
\includegraphics[width=80mm,angle=0]{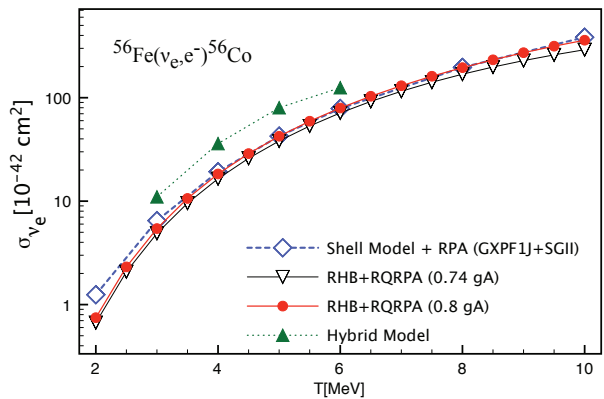}
}
\caption{(Color online) Neutrino - $^{56}$Fe cross sections averaged over 
the Fermi-Dirac distribution. The RHB+RQRPA (DD-ME2) calculations including 
the standard (0.8), and enhanced (0.74) 
quenching factors of the axial-vector coupling $g_A$, are 
compared  to shell model + RPA results \cite{Suz.09}, and those obtained using a hybrid model of 
Ref.~\cite{Kol.01}.
}
\label{Fe56_cs_T_Eph200_Jmax5.eps}
\end{figure}

We have also analyzed cross sections averaged over the neutrino flux described 
 by the Fermi-Dirac spectrum~\cite{Paar.08}.
%
%
Figure \ref{Fe56_cs_T_Eph200_Jmax5.eps} displays the 
flux-averaged cross sections for the reaction $^{56}$Fe($\nu_e,e^-$)$^{56}$Co, 
evaluated at different temperatures in the interval $T= 2 - 10$ MeV, and for
the chemical potential $\alpha$=0.
The RHB+RQRPA (DD-ME2) calculations including the standard (0.8), and enhanced (0.74) 
quenching factors in $g_A$, are 
compared  to the shell model + RPA (GXPF1J + SGII) results \cite{Suz.09}, and those 
obtained using a hybrid model of Ref.~\cite{Kol.01}. The latter model predicts somewhat 
larger cross sections, whereas a very good agreement is found between the results of 
shell model + RPA and RQRPA.  

The theoretical cross sections for the reaction $^{56}$Fe($\nu_e,e^-$)$^{56}$Co can also 
be analyzed in comparison to data from the KARMEN collaboration. The calculated
cross sections are averaged over the neutrino flux described by the Michel spectrum
obtained from muon decay-at-rest (DAR): $f(E_{\nu_e}) = 
 ({96 E_{\nu_e}^2}/{m_{\mu}^4})  \left (  m_{\mu} -2
E_{\nu_e}  \right )$. Taking into account the results obtained in this work with the 
RHB+RQRPA (DD-ME2) - 263 $\times$ 10$^{-42}$cm$^2$ , shell model (GXPF1J) (for $1^+$ transitions) plus the RPA (SGII)
for other multipoles - 259 $\times$ 10$^{-42}$cm$^2$ , as well as results from previous studies that used the RPA with a Landau-Migdal
force - 240 $\times$ 10$^{-42}$cm$^2$~\cite{Kol.01}, the QRPA(SIII) - 352 $\times$ 10$^{-42}$cm$^2$~\cite{Laz.07} and the QRPA based on G-matrix formalism - 173.5 $\times$ 10$^{-42}$cm$^2$~\cite{Che.10}, 
the DAR neutrino-nucleus cross section and its theoretical uncertainty are estimated to be: $<\sigma>_{th}$=(258$\pm$57) $\times$10$^{-42}$cm$^2$. This value is in very good agreement with the data from 
the KARMEN collaboration: $<\sigma>_{exp}$=(256$\pm$108$\pm$43) $\times$ 10$^{-42}$cm$^2$.
We note that the various models used to obtain the theoretical estimate employ different 
effective interactions, and also comprise a wide range of values for the axial-vector coupling, 
from those without quenching~\cite{Laz.07, Che.10} to models that use a quenching factor of 
0.7~\cite{Kol.01}. All theory frameworks, except the QRPA based on G-matrix formalism, favor 
the quenching of the axial-vector coupling constant $g_A$, in accordance to constraints given by 
the experimental data on Gamow-Teller transitions. The implementation of the quenching 
of  $g_A$ in QRPA based on Skyrme functionals~\cite{Laz.07}  would lower
the calculated neutrino-nucleus cross sections and it would further reduce the overall 
theoretical uncertainty in $<\sigma>_{th}$.

In conclusion, the charged current neutrino-nucleus cross sections for $^{54,56}$Fe and $^{58,60}$Ni
have been analyzed by employing models based on the relativistic and Skyrme energy density functionals, 
and the shell model. The theoretical uncertainties in modeling neutrino induced processes 
have been examined by considering the Gamow-Teller transition strength and available data, the multipole decomposition of the calculated cross sections, and cross sections averaged over the Michel flux and 
Fermi-Dirac distribution. It has been shown that various models predict very similar multipole distributions 
of neutrino-nucleus cross sections. The corresponding cross sections averaged over the DAR neutrino spectra
show that the current theoretical uncertainty, despite a variety of models and effective interactions that 
have been used in many studies, is actually smaller than the experimental one, and could be even further 
reduced by constraining the quenching of the axial-vector coupling constant $g_A$ to data for the 
Gamow-Teller transition strength.

\bigskip
\leftline{\bf ACKNOWLEDGMENTS}
\noindent
This work is supported by  MZOS - project 1191005-1010, the Croatian Science Foundation, 
and Grants-in-Aid for Scientific Research (C) 22540290 of the MEXT of Japan.


\begin{thebibliography}{999}

\bibitem{Ath.97} C. Athanassopoulos et al., Phys. Rev. C 55, 2078 (1997).

\bibitem{Bod.94} B. E. Bodmann et al., Phys. Lett. B 332, 251 (1994).

\bibitem{Mas.98} R. Maschuw, Prog. Part. Nucl. Phys. 40, 183 (1998).

\bibitem{Kra.92} D. A. Krakauer et al., Phys. Rev. C 45, 2450 (1992).

\bibitem{Avi.03} F. T. Avignone, L. Chatterjee, Yu. V. Efremenko,
M. Strayer, J. Phys. G 29, 2497 (2003).

\bibitem{Aga.07} N.Yu. Agafonova et al.,  Astron. Phys. 27, 254 (2007).

\bibitem{Zuc.02} P. Zucchelli, Phys. Lett. B 532, 166 (2002).

\bibitem{Vol.04} C. Volpe, J. Phys. G 30, L1 (2004).

\bibitem{Hax.87} W. C. Haxton, Phys. Rev. D 36, 2283 (1987).

\bibitem{Eng.96} J. Engel, E. Kolbe, K. Langanke, and P. Vogel, Phys. Rev. C 54, 2740 (1996).

\bibitem{Hay.00} A. C. Hayes and I. S. Towner, Phys. Rev. C 61, 044603 (2000).

\bibitem{Vol.00} C. Volpe, N. Auerbach, G. Col\`o, T. Suzuki, and N. Van Giai,
Phys. Rev. C 62, 015501 (2000).

\bibitem{Suz.09} T. Suzuki, M. Honma, K. Higashiyama, T. Yoshida, T. Kajino, T. Otsuka, H. Umeda, 
and K. Nomoto, Phys. Rev. C 79, 061603(R) (2009).

\bibitem{Aue.97}  N. Auerbach, N. Van Giai, and O. K. Vorov,
Phys. Rev. C 56, R2368 (1997).

\bibitem{Sin.98} S. K. Singh, N. C. Mukhopadhyay, and E. Oset,
Phys. Rev. C 57, 2687 (1998).

\bibitem{Kol.92} E. Kolbe, K. Langanke, S. Krewald, F.-K. Thielemann,
Nucl. Phys. A 540, 599 (1992).

\bibitem{Jac.99} N. Jachowicz, S. Rombouts, K. Heyde, and J. Ryckebusch, 
Phys. Rev. C 59, 3246 (1999).

\bibitem{Bot.05} A. Botrugno and G. Co', Nucl. Phys. A 761, 200 (2005). 

\bibitem{KLM.99} E. Kolbe, K. Langanke, and G. Mart{\'i}nez-Pinedo, 
Phys. Rev. C 60, 052801 (1999).

\bibitem{Kur.90} T. Kuramoto, M. Fukugita, Y. Kohyama, and K. Kubodera,
Nucl. Phys. A 512, 711 (1990).

\bibitem{Laz.07} R. Lazauskas and C. Volpe, Nucl. Phys. A 792, 219 (2007).

\bibitem{Che.10} M. K. Cheoun, E. Ha, K. S. Kim, and T. Kajino, J. Phys. G 37, 055101 (2010).
                
\bibitem{Sam.11} A. R. Samana, F. Krmpotic, N. Paar, and C. A. Bertulani, Phys. Rev. C 83, 045807 (2011).      
                
\bibitem{Paar.08} N. Paar, D. Vretenar, T. Marketin, and P. Ring, Phys. Rev. C 77, 024608 (2008).

\bibitem{LNVR.05} G. A. Lalazissis, T. Nik\v si\' c, D. Vretenar,
                  and P. Ring, Phys. Rev. C 71, 024312 (2005).
                  
\bibitem{BGG.91} J. F. Berger, M. Girod, and D. Gogny,
                 Comp. Phys. Comm. 63, 365 (1991).
                 
\bibitem{Hon.05} M. Honma, T. Otsuka, T. Mizusaki, M. Hjorth-Jensen and B. A. Brown,
J. of Phys.: Conf. Series 20, 7 (2005). 

\bibitem{Miz.00} T. Mizusaki, RIKEN Accel. Prog. Rep. 33, 14 (2000).                 

\bibitem{Rap.83} J. Rapaport et al., Nucl. Phys. A 410, 371 (1983).

\bibitem{Vet.89} M. C. Wetterli et al., Phys. Rev. C 40, 559 (1989).

\bibitem{Kat.94} S. El-Kateb et al., Phys. Rev. C 49, 3128 (1994).

\bibitem{Wil.95} A. L. Williams et al., Phys. Rev. C 51, 1144 (1995).

\bibitem{Kol.01} E. Kolbe and K. Langanke, Phys. Rev. C 63, 025802 (2001).

    	  	
\end{thebibliography}







\end{document}